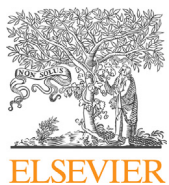

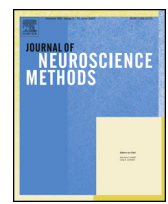

Research Paper

# A novel method linking neural connectivity to behavioral fluctuations: Behavior-regressed connectivity

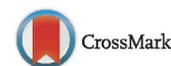

Antony D. Passaro [a,*], Jean M. Vettel [a,b,c], Jonathan McDaniel [d], Vernon Lawhern [a], Piotr J. Franaszczuk [a,e], Stephen M. Gordon [d]

[a] Human Research and Engineering Directorate, U.S. Army Research Laboratory, Aberdeen Proving Ground, MD 21005, USA
[b] University of California, Santa Barbara, CA 93106, USA
[c] University of Pennsylvania, PA 19104, USA
[d] DCS Corporation, Inc., Alexandria, VA 22310, USA
[e] Johns Hopkins University, Baltimore, MD 21205, USA

## HIGHLIGHTS

- Conventional connectivity derives a single connectivity pattern for a behavior.
- We propose Behavior Regressed Connectivity (BRC) to track behavioral fluctuations.
- BRC provides a complimentary understanding to conventional connectivity.
- BRC approach identifies participant-specific differences across a range of behaviors.

## ARTICLE INFO

## ABSTRACT

Background: During an experimental session, behavioral performance fluctuates, yet most neuroimaging analyses of functional connectivity derive a single connectivity pattern. These conventional connectivity approaches assume that since the underlying behavior of the task remains constant, the connectivity pattern is also constant.

New method: We introduce a novel method, behavior-regressed connectivity (BRC), to directly examine behavioral fluctuations within an experimental session and capture their relationship to changes in functional connectivity. This method employs the weighted phase lag index (WPLI) applied to a window of trials with a weighting function. Using two datasets, the BRC results are compared to conventional connectivity results during two time windows: the one second before stimulus onset to identify predictive relationships, and the one second after onset to capture task-dependent relationships.

Results: In both tasks, we replicate the expected results for the conventional connectivity analysis, and extend our understanding of the brain-behavior relationship using the BRC analysis, demonstrating subject-specific BRC maps that correspond to both positive and negative relationships with behavior.

Comparison with Existing Method(s): Conventional connectivity analyses assume a consistent relationship between behaviors and functional connectivity, but the BRC method examines performance variability within an experimental session to understand dynamic connectivity and transient behavior.

Conclusion: The BRC approach examines connectivity as it covaries with behavior to complement the knowledge of underlying neural activity derived from conventional connectivity analyses. Within this framework, BRC may be implemented for the purpose of understanding performance variability both within and between participants.

Published by Elsevier B.V. This is an open access article under the CC BY license (http://creativecommons.org/licenses/by/4.0/).

* Corresponding author at: Human Research and Engineering Directorate, U.S. Army Research Laboratory, 459 Mulberry Point Road, Aberdeen Proving Ground, MD 21005, USA.
  E-mail addresses: Antony.Passaro@gmail.com (A.D. Passaro), jean.m.vettel.civ@mail.mil (J.M. Vettel), jmcdaniel@dcscorp.com (J. McDaniel), Vernon.j.lawhern.civ@mail.mil (V. Lawhern), Piotr.j.franaszczuk.civ@mail.mil (P.J. Franaszczuk), sgordon@dcscorp.com (S.M. Gordon).

# 1. Introduction

Behavioral performance fluctuates both within and across tasks and individuals, yet analyses typically derive a single pattern of connectivity across individuals for a single task (e.g., Stam et al., 2007; Jin et al., 2012; Chen et al., 2013). Functional connectivity metrics describe either the direct or indirect relationship within the frequency domain between two signals recorded from the brain, and previous reviews document their utility for examining brain-behavior relationships (e.g., Gourévitch et al., 2006; Wendling et al., 2009; Friston, 2011; He et al., 2011; Sakkalis, 2011). Summarizing a functional connectivity pattern as a single, task-dependent configuration often produces an over-simplified representation of the corresponding behavioral processes. Neuroimaging studies have recently demonstrated that these connectivity patterns fluctuate in time as the task progresses (Monto et al., 2008; Stefanics et al., 2010; Sadaghiani et al., 2015; Bassett et al., 2011), but this research has largely ignored components of time-evolving behavioral performance. As a result, both within and across participants, conventional connectivity analyses often yield a small concordance of connectivity patterns across subjects (e.g., Fingelkurts et al., 2007; Gruber and Müller, 2005). Therefore, in order to track dynamic fluctuations in behavior over an experimental session, connectivity methods are needed that capture this temporal variability for a given task.

The variability in both neuroanatomy and neural function has been shown to be expansive between individuals (Frost and Goebel, 2012; Mueller et al., 2013; Sugiura et al., 2007), and it cannot be entirely explained by other sources such as demographics, performance, and strategy (Miller et al., 2012). Critically, this variability is often treated as noise for most group-level neuroimaging studies rather than as an informative component about the individual in particular and understanding the human brain in general. Additionally, this variability may be driven in large-part by state changes over time, such as changes in levels of vigilance, fatigue, stress, and so forth. In theory, individual relationships between structure and function might produce consistent results, however, state perturbations may preclude that possibility as different brains regions are believed to underlie specific states (e.g., Cook et al., 2007; Dedovic et al., 2009; Olbrich et al., 2009). An increasing number of studies have begun to focus on participant-specific analyses to further elucidate neuro-functional differences (Buckelmüller et al., 2006; Drew and Vogel, 2008; Lim et al., 2013; Dong et al., 2015) and in some cases, to lay the foundation for improvement of brain-computer interfaces (BCIs) (Guger et al., 2000; Marathe et al., 2015; Mensh et al., 2004; Thomas et al., 2009; Wu et al., 2013).

A novel method is introduced in this paper, behavior-regressed connectivity (BRC), to directly analyze the fluctuations in behavior and elucidate how this performance variability relates to functional connectivity. The BRC approach requires a time-series of a behavioral measure and the associated time-series of connectivity. Classically, a behavioral measure is the response time or accuracy for sequential trials in an experiment, but for BRC, the time-series can be derived from any task-relevant behavioral metric that can be sampled throughout the experimental session. To compute the associated time-series of connectivity, the proposed approach uses a weighting function applied to a shifting window of trials to mitigate the low signal-to-noise ratio (SNR) of neuroimaging data and compute a stable estimate of connectivity for each window. The sequential windows of trials produce a time-series of neural dynamics synchronized with a time-evolving behavioral time-series. The BRC method then utilizes these two time-series in a linear regression of behavior to connectivity and identifies a set of connections that is relevant for a particular behavior of interest and largely immune to other fluctuations in the neural signal. Due to the use of a shifting window of trials across the experimen-tal session, a critical issue is the tradeoff between the number of trials needed for a stable connectivity estimate and the temporal resolution of the overall analysis. The weighting function also plays a role in the stability estimate and further serves as a connectivity smoothing parameter. These two parameter changes are examined to determine the extent to which the shifting window of trials and weighting function influence the BRC results.

We investigated the versatility of the BRC method by applying it to two tasks that capture two different classes of behaviors: goal-directed behaviors, where a participant is incentivized for their performance by the demand characteristics of the experiment, and non-goal-directed behaviors, where unattended behaviors during a task reveal idiosyncratic tendencies of a particular participant and do not directly relate to the outcome of the task. Data from the first task examines goal-directed behaviors, where the participants actively monitor their score relative to a virtual competitor in a target identification task. The second dataset investigates a non-goal-directed behavior, where the participant is asked to press a button at a self-paced interval while the analysis focuses on how long the button is depressed. We examine two time windows: the one second before stimulus/motor onset to identify predictive relationships, and the one second after onset to capture task-dependent relationships. Combined, our novel method reveals reliable relationships between connectivity and behavior in a dynamic subject-specific manner, and it lays the foundation for extended analyses on behavior prediction and BCI applications.

# 2. Materials and methods

## 2.1. Overview

Functional connectivity measures identify the temporal correlation among neurophysiological events (Friston, 1994). A sub-class of these measures, known as phase synchronization (PS) measures, analyzes differences in the phases of two signals and looks for patterns of consistent activity. This is, in effect, measuring whether there is a consistent time delay between activities in a specific frequency band across the two events (Rosenblum et al., 1996). To produce this measurement, phase-based methods must be computed over multiple instances. This, in turn, hampers the experimenter's ability to analyze trial-by-trial fluctuations in connectivity and, thus, link these fluctuations with observed behavior. To overcome this problem, one option would be to sort the trials by behavior. This approach, however, could distort any temporal, tonic state effects. An alternative approach is described in this paper where the behavior-connectivity relationship is explored using a windowing function. As depicted in Fig. 1, BRC computes the relationship between a time-varying behavioral measure, such as trial-by-trial reaction time, and a specific connectivity estimate.

In this section we provide an overview of our approach. Here, and throughout the paper, we use the WPLI (Vinck et al., 2011) as our sample PS method. We chose this measure because we felt that it best represented the performance of the general class of PS methods. In addition, our own empirical tests, as well as prior research (Vindiola et al., 2014; Gordon et al., 2013), suggests that the WPLI is similar not only in design but also in performance to other PS techniques, such as those described by Stam et al. (2007), Lachaux et al. (1999) and Nolte et al. (2004). After introducing the computation of the WPLI, we then describe the modification to that function in order to produce a windowed estimate before relating that estimate to behavior. This modification can be achieved by replacing the traditional ensemble averaging step by a windowing function with time-varying weight and then performing a sliding estimate for each subsequent trial. This produces a set of trial-by-trial measurements for both behavior and connectivity. We then regress behavior against the connectivity estimates for each scalp electrode pair. In this section, we also describe our approach to analyze the effects of using different windowing functions and window sizes, and we describe the statistical tests we perform. Here, we introduce an alternative to the conventional resampling approach used within the connectivity literature. Finally, we describe two experimental data sets that we use to validate and compare our method.

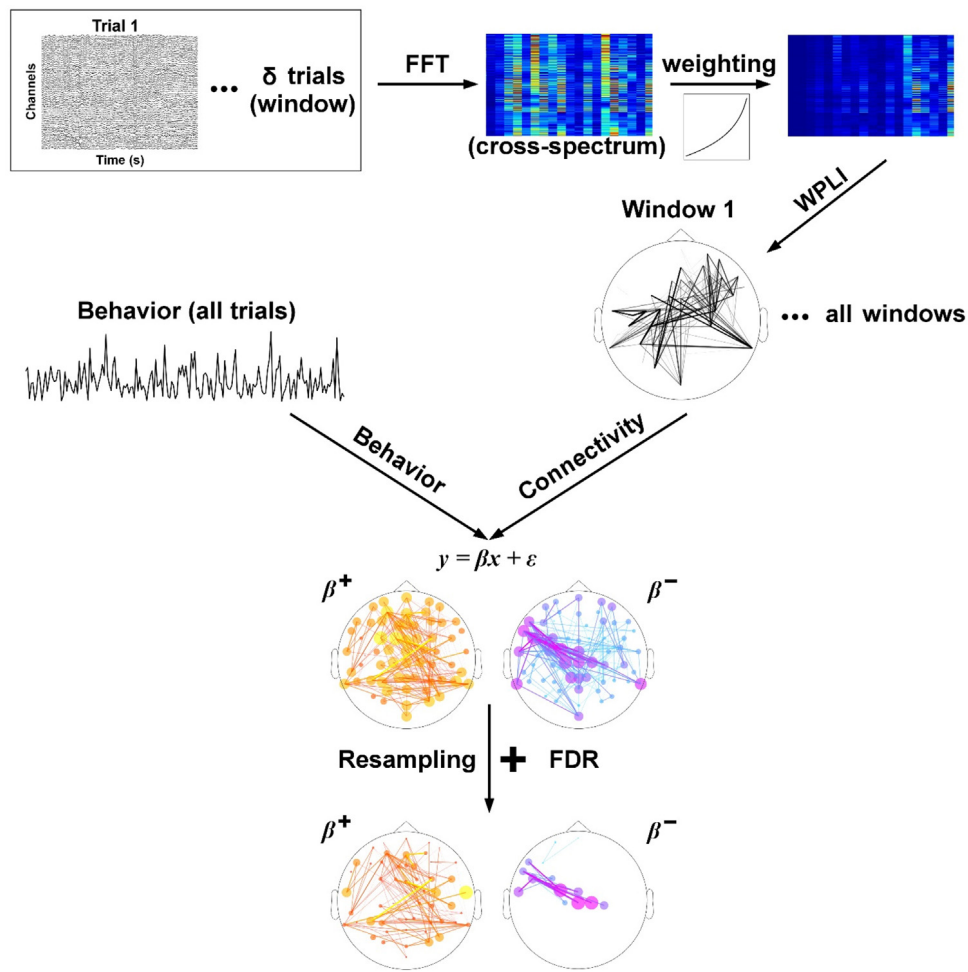

**Fig. 1.** Structure of analysis. Trials of data that cover periods of interest across all channels are grouped into a window (top-left). A spectral decomposition yielding the cross-spectrum followed by a weighting function is applied to the time-series data within the first window of trials (top-right). The WPLI connectivity is computed on the weighted window of trials, and this process is repeated for all windows. The trial-by-trial behavioral time-series (shifted by the window length, δ) is regressed for each connection across all windows of trials (middle) to produce the BRC map of positive coefficients (orange) and negative coefficients (magenta). The results are compared against behavior-shuffled surrogate data using a permutation test (10,000 iterations) and corrected for multiple comparisons across sensor pairs using an FDR correction (bottom). (For interpretation of the references to colour in this figure legend, the reader is referred to the web version of this article.)

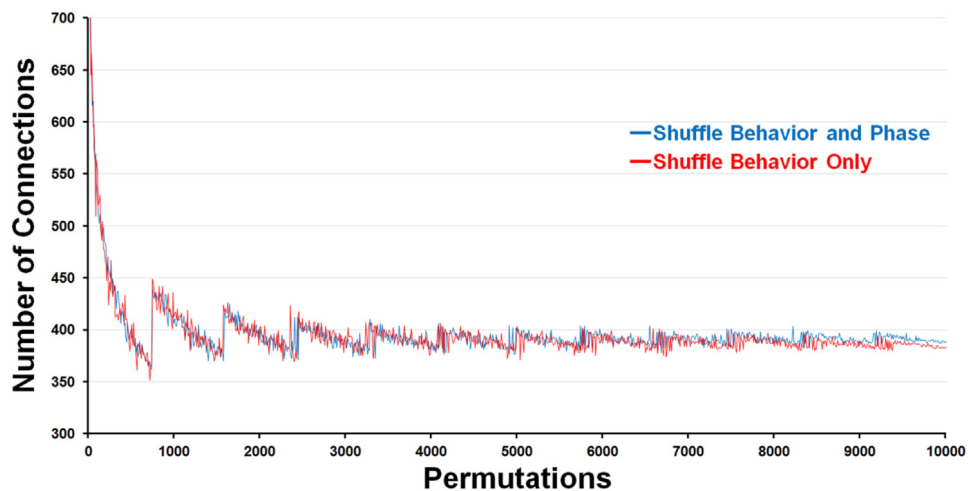

**Fig. 2.** Comparison of resampling approaches. The amounts of statistically significant connections resulting from the BRC approach after FDR correction are illustrated for permutations from 1 to 10,000. The red line corresponds to the BRC results when shuffling only behavior while keeping the original WPLI intact. The blue line corresponds to the BRC results when shuffling both behavior and phase within the EEG data before recomputing the regression. (For interpretation of the references to colour in this figure legend, the reader is referred to the web version of this article.)

### 2.2. Behavior-regressed connectivity

The initial step for computing BRC is to obtain a time-varying behavior (1), where N is a trial index.

$$\text{Behavior}(N) = \text{behavior of the N}^{th}\text{ trial} \tag{1}$$

Next, a modified functional connectivity metric must be computed. In this paper, we utilize the Weighted Phase Lag Index (WPLI) (2), originally introduced by (Vinck et al., 2011) is utilized. The WPLI uses the imaginary component of the cross-spectra in the numerator and denominator to lessen the effect of background power estimates. It has also been shown to be more immune to volume conduction and measurement noise while exhibiting increased sensitivity to phase interactions between signals (Vinck et al., 2011; Vindiola et al., 2014).

$$WPLI_{jk} = \frac{|E\left\{Imag\left(Z_{jk}\right)\right\}|}{E\left\{|Imag\left(Z_{jk}\right)|\right\}} \tag{2}$$

$$Z_{jk} = X_j X_k^* \tag{3}$$

$$X(f) = \int_{-\infty}^{\infty} win(t)x(t)e^{i2\pi ft}dt \tag{4}$$

Where $Z_{jk}$ is the cross-spectrum for electrodes j and k, as defined in (3). The notation $X^*$ denotes the complex conjugate and the function $Imag()$ returns only the imaginary component of the cross-spectrum. In (4), the term $win(t)$ is a windowing function that is often applied prior to the computation of the time-frequency transform in order to produce a smoother estimation by reducing the edge effects that occur when applying a Fourier transform to a discrete, finite signal. The windowing function as defined in this method encompasses a collection of consecutive trials rather than time points, across which the subsequent connectivity metric is computed. In this application of the BRC method, a multi-taper approach was utilized that involved a discrete prolate spheroidal sequence (dpss) taper at 10 Hz with a smoothing width of 3 Hz to produce a complex cross spectrum (Percival and Walden, 1998). $X_j$ is the frequency transformation of the original time-series, $x_j(t)$, which is computed for each trial $n$. $M$ is the total number of trials. The term $E\{\}$ is the expectation operator, which is typically approximated using the form shown in (5).

$$E\left\{Z\right\} \approx \sum_{n=1}^{M} w(n)Z(n) \tag{5}$$

$$w(n) = \frac{1}{M} n = 1 \ldots M \tag{6}$$

WPLI is typically calculated using all trials of interest, and while a certain amount of averaging across trials is required in order to accurately assess phase differences, this static approach ignores transient changes in functional connectivity. Therefore, a shifting window approach is used whereby the WPLI is calculated within small groups of trials (δ), with δ−1 overlap between groups of trials. For a trial N, the associated WPLI value is computed from the δ−1 trials preceding N and trial N itself. This generates a connectivity value based on δ that is associated with each trial N and the corresponding behavior for that trial.

Within each window of trials, an exponential weighting is applied to place greater emphasis on the more recent trials that are closer in time to the matched behavior value and less emphasis to those trials further in time. The cross spectrum values within each window of trials for δ are weighted using (7), which replaces (6).

$$w(n, N) = e^{-(N-n)\frac{1}{\tau}} \tag{7}$$

Using this approach, a weighted estimate of the WPLI is obtained for the trials up to, and including, trial N, across a window length of trials (δ). The WPLI for δ can be paired with the behavioral measures (1) to support standard regression analysis (8).

$$WPLI = \beta * Behavior + \varepsilon \tag{8}$$

Where β and ε are regression coefficients.

### 2.3. Time-evolving connectivity

For each task, the WPLI is computed for the alpha range (7–13 Hz) of a 1 s period using 64 channels of scalp data. Two different 1 s epochs are analyzed. To examine predictive brain to behavior relationships, the one second period (−1 to 0 s) preceding stimulus onset (motor onset in the case of finger-tapping) is studied, and then a one second period (0 to 1s) following stimulus/motor onset is analyzed to study task-dependent brain to behavior relationships. The data in each window is spectrally decomposed for all trials within that window to derive the cross-spectrum of the alpha band, due to that band's role in modulating behavior across a variety of tasks (e.g., Van Dijk et al., 2008; Jokisch and Jensen, 2007; Haegens et al., 2011). The bivariate WPLI implementation of the BRC method is described here in which the upper diagonal of all 4096 signal pairs represents the total number of unique signal pairs for this measure for a total of 2016 unique connections [(4096 − 64)/2].

### 2.4. Weighting functions and size of trial windows

The WPLI is an average measure that requires a shifting window to incorporate information from multiple trials, and as a result, this measure and its relationship to behavior will fluctuate depending on the weighting function and the window length of trials. We assessed four weighting functions, including the exponential weighting function (presented in Eq. (7)), a boxcar, a logarithmic, and half of a Hanning function (Blackman and Tukey, 1959). All weighting functions are illustrated in Figs. 3 and 4. To assess the effect of the number of trials within a window (δ), a variable number of trials were used, ranging from 5 trials up to half of the length of total trials. Using only half of each dataset in the BRC approach was required to control for an equal number of degrees of freedom across window sizes, and we used half of the shorter dataset to ensure equivalence between the two tasks (half of finger-tapping dataset was 295 trials). To identify the optimal window length separately for each dataset, the value was determined when the number of added/removed statistically significant connections varied by less than 10% across successive window sizes and the value reached near an asymptote. As a complementary measure, the maximum absolute regression coefficient was converted to a t-statistic by dividing by the standard error of the coefficient, and this t-statistic is also plotted to examine its stability in relation to changes in statistically significant connections.

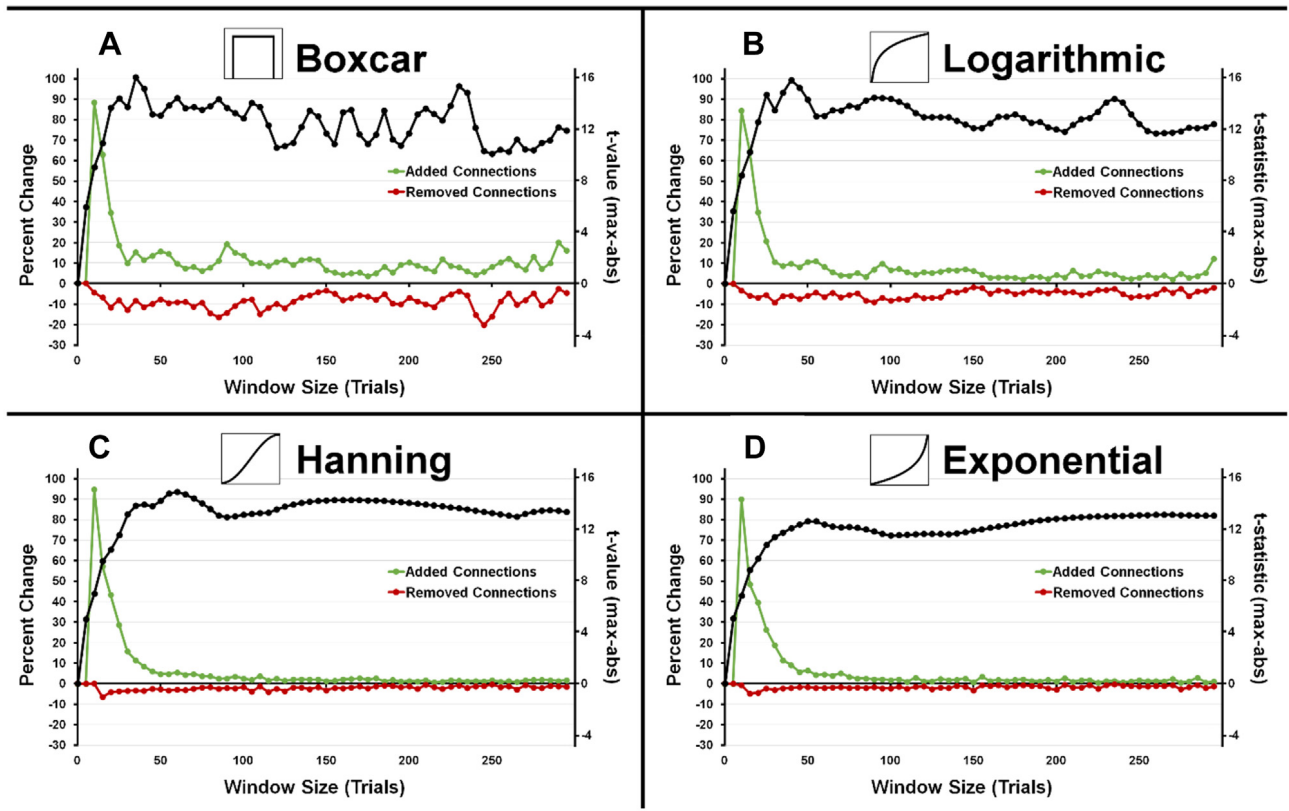

**Fig. 3.** Comparison of window sizes of trials and weighting functions for the target identification task. Results from the BRC analysis of the target identification task as described by the number of statistically significant connections added (green points) and removed (red points) and the maximum absolute regression coefficient converted to a *t*-statistic (black points) are illustrated for each window size from 5 to 295 trials for each of the four tested weighting functions. A. BRC results across window sizes for the boxcar weighting function. B. BRC results for the logarithmic weighting. C. BRC results for half of a traditional half-Hanning function. D. BRC results for the exponential weighting function, the preferred weighting for the BRC method. (For interpretation of the references to colour in this figure legend, the reader is referred to the web version of this article.)

## 2.5. Conventional connectivity

The conventional approach to computing connectivity identifies connections that are derived from a task or a condition across the full duration of the experimental session (all trials) without accounting for behavioral fluctuations. To assess how it differs from the BRC approach, the conventional connectivity analysis was computed for each task and tested for statistical significance without the use of regression. The WPLI was computed using data epoched around the 1 s period following stimulus/motor onset and the cross spectrum was derived from the FFT focusing on the alpha band (7–13 Hz). Importantly, the WPLI was computed across all available trials, rather than a subset of trials, to yield a single WPLI value for each connection within each task.

## 2.6. Statistical significance

For each unique connection, i.e. pair of electrodes $j$ and $k$, a linear regression was computed across trials using (8). The sign and strength of the regression coefficient, $\beta$, is associated with each signal pair to determine if a positive or negative relationship with behavior exists. The regression coefficient was converted into a *t*-statistic by dividing by its standard error, and then it was used to determine the statistical significance associated with each signal pair by comparing the range of values derived from the experimental data to the values derived from a surrogate dataset. For each dataset, a separate set of surrogate data was computed by phase-shuffling the EEG data as described by Theiler et al. (1992). The phase is shuffled across frequencies within each trial for each channel, which ensures that the power of the original EEG signal is left intact while the phase is shuffled. Likewise, the behavioral values are shuffled over all trials to eliminate the structure of the behavioral response as well. The weighting function was then applied to each shifting window of trials for the cross spectrum of the phase-shuffled data and the resulting WPLI was calculated. The *t*-statistic representation of the regression coefficient for each signal pair was computed across all trials using the phase- and behavior-shuffled data for each permutation (10,000 times) to create a distribution of values against which to test the significance of the original experimental datasets. The number of permutations was chosen based on results illustrated in Fig. 2, where the estimate of the number of significant connections appears to stabilize after approximately 7000 to 10,000 permutations.

The analysis treats each of the signal pairs independently, and the linear regression with behavior was also performed separately for each pair; therefore, to correct for multiple comparisons, a mod-

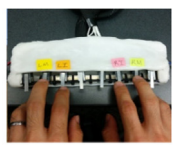

**Finger Tapping**

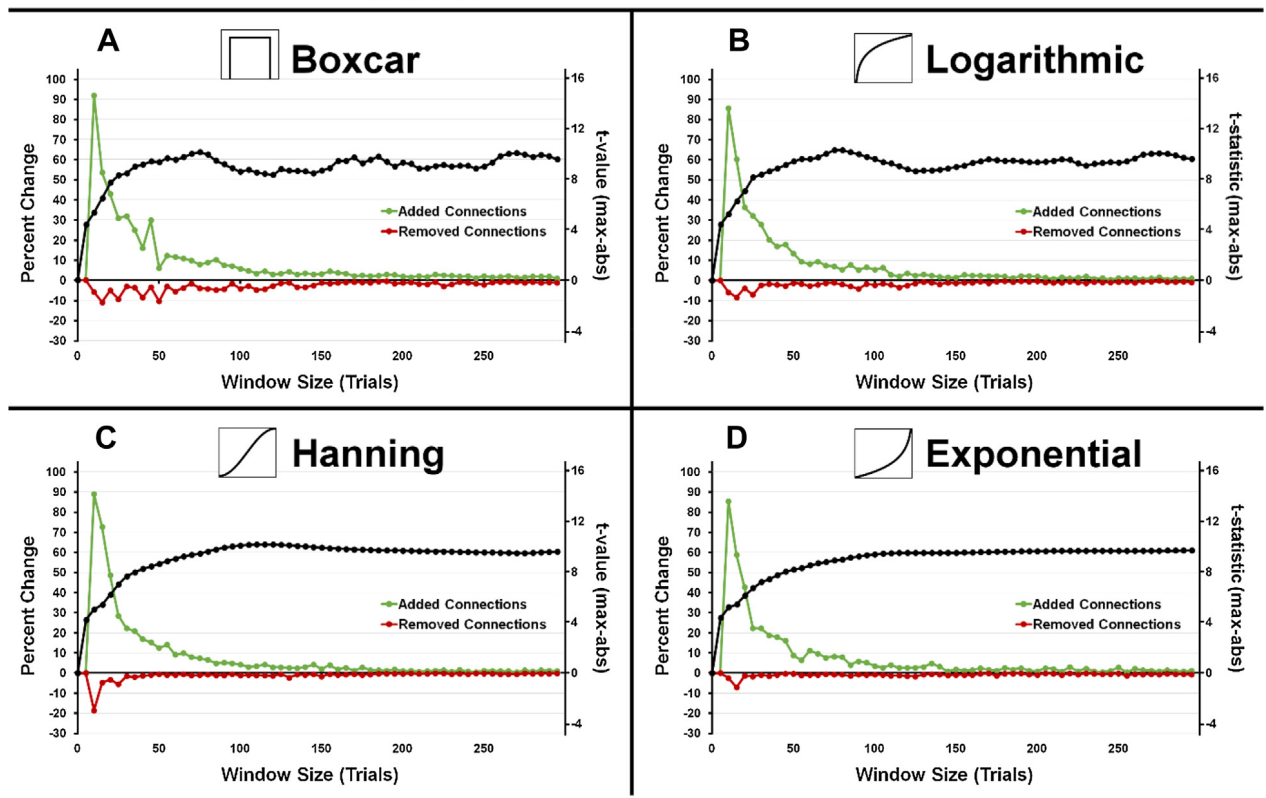

**Fig. 4.** Comparison of window sizes of trials and weighting functions for the finger-tapping task. Results from the BRC analysis of the finger-tapping task as described by the number of statistically significant connections added (green points) and removed (red points) and the maximum absolute regression coefficient converted to a *t-statistic* (black points) are illustrated for each window size from 5 to 295 trials for each of the four tested weighting functions. A. BRC results across window sizes for the boxcar weighting function. B. BRC results for the logarithmic weighting. C. BRC results for half of a traditional Hanning function. D. BRC results for the exponential weighting function, the preferred weighting for the BRC method. (For interpretation of the references to colour in this figure legend, the reader is referred to the web version of this article.)

ification of the false discovery rate (FDR) was applied as described by Benjamini and Yekutieli (2001) that is accurate for all independent tests and those that are positively correlated. A Bonferroni correction across so many signal pairs has the potential to be too stringent, whereas a FDR correction provides more control of the Type I error rate. The signal pairs that pass the FDR correction are reported in the results.

Since no continuous behavior is required for the conventional connectivity analysis, the statistical significance was determined by phase shuffling the functional connectivity values. A non-parametric computation of connectivity followed by the FDR correction was utilized in the conventional connectivity analysis to directly match the BRC approach, even though it is possible to utilize a parametric computation of conventional connectivity due to the availability of all trials rather than just a subset within a window.

### 2.6.1. Resampling alternative

Since both the signal phase and behavior are shuffled in the BRC method, the resampling approach is computationally intensive to perform 10,000 permutations. An alternative resampling approach was therefore investigated that only shuffled the behavior (preserving the original connectivity) in order to destroy the trial-by-trial behavior-connectivity relationship and save computational time. Using the target identification dataset ($\delta = 40$, M = 576 total trials), results showed that this alternative approach reduced the resampling time by a factor of approximately 575. Results comparing this alternative resampling approach to the traditional phase-and-behavior shuffling approach are illustrated in Fig. 2. The number of statistically significant connections resulting from the BRC analysis appears to overlap across the two approaches regardless of the number of permutations. The source of the sawtooth structure is a direct result of the FDR method used to correct for multiple comparisons as the critical p-value jumps once the distribution of p-values significantly changes, which is an indication that the resampling procedure has not yet converged to a stable estimator. This structure in the resampling data diminishes substantially as it approaches the 10,000th iteration, where the estimate of the number of significant connections across both resampling approaches changes by less than 10 as the number of permutations continues to increase. For the analyses exploring the effects of weighting functions and window lengths of trials, this computationally simpler approach was used given the large number of comparisons performed. Moreover, empirical tests comparing both resampling approaches demonstrated that the BRC results between the two methods differed very little and that the overall trends were virtually identical. Therefore, it is recommended to apply the alternative resampling approach that shuffles behavior only (and not phase values) when assessing the statistical significance of the BRC results.

## 2.7. Experimental datasets

The BRC approach is applied to a different participant from two diverse datasets. The BRC analysis of the first dataset examines the connectivity linked with goal-directed behaviors, while the second investigates connectivity related to nuanced changes in unattended, idiosyncratic behaviors. Both datasets were collected in accordance with IRB requirements (32 CFR 219 and DoDI 3216.02). For each experimental paradigm, the data was preprocessed using the Fieldtrip toolbox (Oostenveld et al., 2010) and the parameters of this preprocessing are described in detail for both datasets in the following paragraphs. A specific behavior and time-period were defined for the BRC analysis of each dataset as well.

### 2.7.1. Dataset 1, target identification

A participant competed online with a computer-based competitor in a 30-min target identification task. The participant was driven around a simulated 3D environment, and he was required to discriminate human entities as either a non-target (an image of a man) or a target (an image of a man with a gun) and tables as either a non-target (an image of a table with a clear view under it) or a target (an image of a table with an obstructed view of the space under it). Stimuli were presented for 1 s with an inter-stimulus interval (ISI) of $2 \pm 0.5$ s. The participant responded by pushing a button with the left or right hand to indicate the target type depending on the block while hands were counter-balanced across blocks. A score was associated with each response such that a fast correct response resulted in a high positive score while a slow correct response gave a low positive score. Conversely, a fast incorrect response produced a large negative score and a slow incorrect response produced a small negative score. These scores were cumulative over time to produce a running score across trials. The participant was presented with 300 human entities (150 targets) and 300 tables (150 targets), and he was requested to adopt a strategy to maximize his score to beat the score of the computer-based competitor. Unbeknownst to the participant, the computer-based competitor's score followed a predefined trajectory with respect to the participant's score to ensure that there were extended timeframes when the participant was winning and losing. Consequently, the competitor score was dependent on how well the participant performed in the last five trials, but we added noise to the scoring procedure to make the relationship less obvious to the participant.

EEG data were acquired using a 64-channel BioSemi system (Amsterdam, The Netherlands) sampled at 256 Hz and referenced online to the Common Mode Sense electrode, then re-referenced to the average of the mastoid electrodes offline. The EEG data were then decomposed using an Independent Component Analysis (ICA) (Makeig et al., 1996). Components were visually inspected and those containing artifact signals were identified and subtracted from the original signal. The remaining bad data segments and channels were identified by visual inspection and removed and interpolated, respectively. In this dataset, the behavioral time-series was derived from the difference between the game score awarded for the participant's response and the score of the computer-based competitor. Both connectivity analyses consisted of 576 trials that were time-locked to the target onset.

### 2.7.2. Dataset 2, finger-tapping

A participant was asked to stare at a fixation cross and perform a self-paced finger-tapping task using a metal lever on a homemade button box. The task consisted of 16 five minute blocks, each of which was preceded by instructions informing the participant which finger (index or middle) and hand (alternating across blocks) to use for that particular block. The participant was asked to pace their button presses approximately 5–8 s apart. There were four total blocks for each hand and finger combination.

EEG data were acquired using a BioSemi 256-channel system with a sampling rate of 1024 Hz and referenced online to the Common Mode Sense electrode. The data was then re-referenced to an average of two mastoid electrodes before down-sampled to 256 Hz. Artifact removal processing was performed using the same procedure as in Dataset 1. The functional connectivity measure was computed on a 64-channel subset designed around the classic 10–20 layout. In this dataset, the behavioral time-series was derived from the button-press duration, which is the period of time during which a button was depressed for each button press. Both connectivity analyses were computed for 812 trials that were time-locked to the button-press and combined across sequential blocks.

## 3. Results

The BRC approach was applied to a different participant from two different datasets. In the first dataset, the BRC analysis investigates the relationship between fluctuations in goal-directed performance in a competitive, target identification task and the underlying connectivity. In the second dataset, the BRC method is applied to investigate the relationship between connectivity and the duration that a participant depressed a button in a self-paced finger-tapping task, a task-irrelevant behavior in the experiment. The WPLI functional connectivity measure was used to compute activity between all sensor pairs for the alpha band (7–13 Hz). First, the parameter changes for the number of trials within the window and the weighting function were examined to determine their influence on the BRC results. Then, using parameters derived from this analysis, results for the BRC approach for the one second before stimulus/motor onset was derived to examine predictive brain-to-behavior relationships ($-1-0$ s) and the one second after onset ($0-1$ s) to examine task-dependent relationships. Results for the conventional connectivity approach are also reported for the same two time periods of interest.

### 3.1. The effects of window size and weighting function

The BRC results for different window lengths of trials and weighting functions are illustrated in Figs. 3 & 4 for both tasks during the 1 s following stimulus/motor onset. Each subplot (A–D) shows the results for one of the four weighting functions. To examine the effect of the number of trials within a window, the number of statistically significant connections added (green points) and the number of connections removed (red points) are plotted as a percentage change relative to the previous window size. The maximum absolute regression coefficients across all significant connections converted to a t-statistic (black points) are plotted for each window size from 5 to 295 trials ($M = 576$ for target identification and $M = 812$ for finger-tapping). In the target identification task (Fig. 3), the boxcar weighting produced the most volatile results across window sizes with no apparent convergence, while the exponential weighting produced the most stable BRC results. Using the guideline of a 10% change in total connections added or removed (red and green dots), the exponential weighting fell below that threshold at 50 trials, which also coincided a rough peak of the maximum absolute t-statistic (black dots). A change in the added/removed connections didn't fall below this threshold until much larger window sizes for the boxcar and logarithmic weightings and the half-Hanning weighting exhibited two peaks around the 50 trial window size. Results from the finger-tapping task (Fig. 4) demonstrated a similar trend with the boxcar weighting producing the most volatile BRC across window sizes and the exponential weighting producing the most stable BRC. For this task, the total number of added/removed connections fell below the 10% threshold at a window size of 70 trials, which was smaller compared to the other 3 weighting functions. Therefore, the exponential weighting function was used for the BRC analysis across both tasks, utilizing a window size of 50 trials for the target identification task and 70 trials for the finger-tapping task. The parameter convergence illustrated in these analyses across two different tasks suggests that the observed relationships are robust for each task, individual, and importantly, the cognitive state of the individual during the experiment.

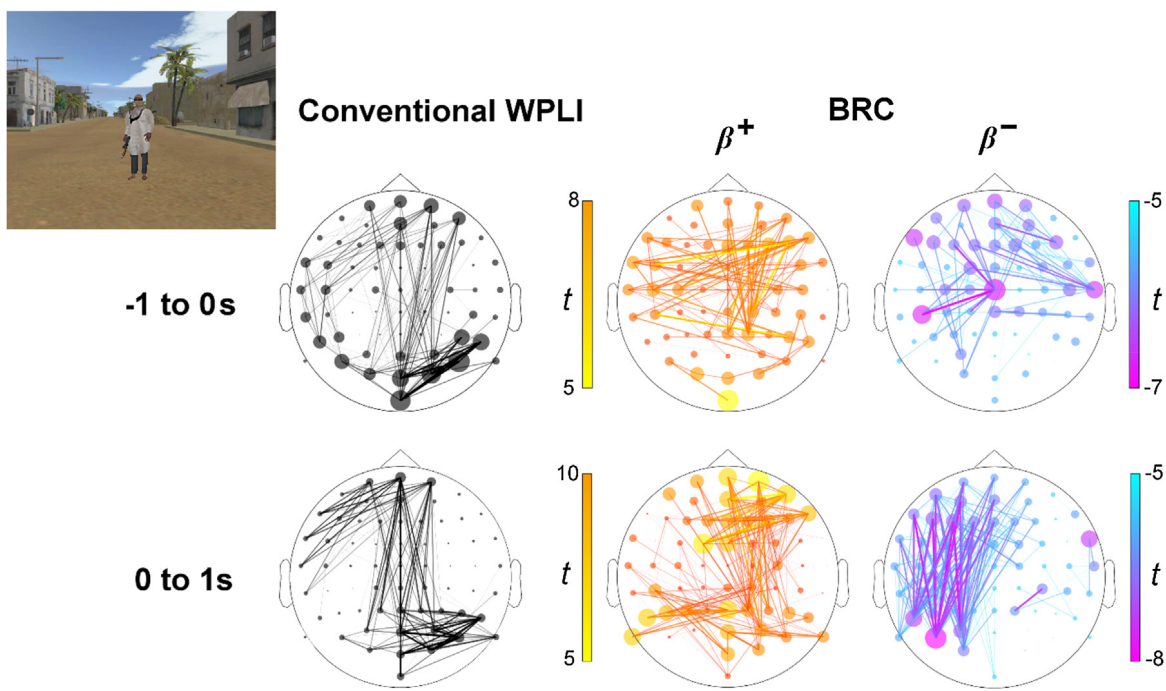

**Fig. 5.** Target identification results. Conventional WPLI across all trials for the target identification task illustrated in black for both the −1 to 0 s period and the 0–1 s period in the first column. Results from the BRC analysis are displayed as connections representing a positive relationship to behavior as illustrated in orange/yellow (second column) and blue/magenta for negative relationship connections (third column). Conventional connectivity patterns are consistent across time periods (top to bottom), while BRC results differentiate between time periods.

### 3.2. Target identification connectivity results

As shown in Fig. 5, the conventional connectivity analysis across all trials for the target identification task produces a similar connectivity pattern across both the time period before stimulus onset (−1–0 s) and the period following stimulus onset (0–1 s). These patterns consist primarily of long WPLI connections over midline sensors, a pattern in accordance with previous work that used EEG to study target identification (Mulert et al., 2004; Busch et al., 2009). The BRC results for the −1 to 0 s period (top-right) produces a different pattern of connections than the conventional connectivity analysis. BRC identifies connections in channels located over right frontal cortex that exhibit a positive relationship, where increased connection strength links to increased performance. BRC also identifies a diffuse central pattern for connections exhibiting a negative relationship, where increased connection strength links to decreased performance. These patterns changed for the BRC analysis of the 1 s period that followed stimulus onset (0–1 s). More positive relationship connections appeared in the front right and posteriorly while negative relationship connections dominated the left side. Importantly, the BRC results illustrate a different set of connections than the conventional connectivity approach which reflects a fundamental difference between these two approaches and exemplifies their complementary nature. Once the connectivity is directly linked to a trial-by-trial fluctuation in behavior rather than lumped together to generally describe a behavior for the conventional connectivity approach, a completely different network emerges.

### 3.3. Finger-tapping connectivity results

Results for the finger-tapping task are shown in Fig. 6. In the conventional connectivity analysis, WPLI activity within the alpha band was observed for sensor pairs over motor and premotor regions as expected from previous literature (Muthukumaraswamy et al., 2004; Stavrinou et al., 2007; Herz et al., 2012). This pattern of connectivity was consistent across both time periods: the 1 s period before the button press (finger completely depressed the lever) and the 1 s period after the button press. Again, the BRC results tell a complementary story. Only a few short range frontal connections and a few long range anterior-posterior connections capture a predictive, positive relationship between brain activity and button press duration (−1–0 s). The BRC results for the period following the button press produce substantially more positive relationship connections over posterior left to middle sensors as well as some short range connections over front left sensors. Only a few negative relationship connections were found for the 1 s period following the button press and primarily covered front left sensors. Once again, the BRC pattern that tracks behavioral fluctuations illustrates a complimentary set of connections when compared to the conventional connectivity approach.

## 4. Discussion

Much of the neuroimaging literature links behavior to neural signals indirectly by contrasting two or more conditions to discern global neural representations for a particular behavior across participants, or researchers sort trials into quartiles to understand brain activity during trials with similar response profiles. Our BRC method diverges from this approach and directly links behavior to connectivity in a participant-specific, time-varying manner to identify robust relationships. These relationships are believed to dynamically change depending on the cognitive state of the individual. In the BRC method, we preserve the order in which the trials occurred to directly examine how variability of performance within session relates to dynamic changes in functional connectivity. Furthermore, this method can be used to identify predictive relationships by focusing on a time period of observed connectivity that precedes the behavior, such as the −1 to 0 s period examined in these two tasks.

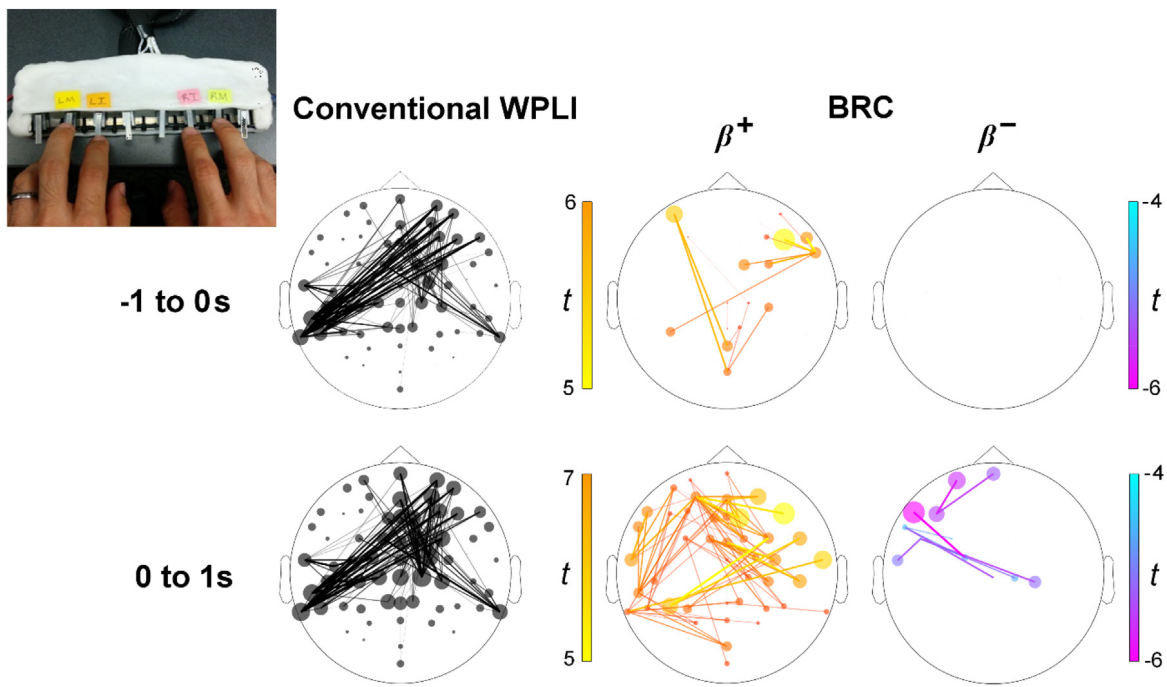

**Fig. 6.** Finger-tapping BRC results. Conventional WPLI across all trials for the finger-tapping task illustrated in black for both the −1 to 0 s period and the 0–1 s period in the first column. Results from the BRC analysis are displayed as connections representing a positive relationship to behavior as illustrated in orange/yellow (second column) and blue/magenta for negative relationship connections (third column). Conventional connectivity patterns are consistent across time periods (top to bottom), while BRC results differentiate between time periods.

The application of this connectivity approach has been demonstrated across two distinct tasks and behaviors involving target identification and self-paced finger-tapping. These behaviors represent examples of both overt, goal-directed behavior (e.g., score in a competitive target identification task) and unattended, non-goal-directed behavior (e.g., button-press duration during finger-tapping). In both cases, the expected results for the conventional connectivity analyses were replicated, and our understanding of brain-behavior relationships was extended when using the BRC method to identify a complementary subset of connections. The diversity of tasks and behaviors for which the BRC approach may be applied is limited only by the frequency at which the behavior is recorded during a task. Any task that allows for the recording of a trial-by-trial or time-by-time behavior in concert with a neuroimaging recording can be analyzed with the BRC method to reveal how behavioral fluctuations are captured in time-evolving connectivity.

### 4.1. Methodological parameters

Several of the parameters used in this application of the BRC method can be adapted to apply the method to alternative research questions. The two main parameters that we explored in this paper were window length of trials and the weighting function within that window. The replicability of this method under the influence of these two critical parameters demonstrates robust findings under the following assumptions: both behavior and connectivity fluctuate as a direct result of state changes for the individual. We also focused on two different types of behavior: a goal-directed behavior and an unattended non-goal directed behavior. These parameters are fundamental to the novel BRC method that we introduce; however, many alternative approaches and parameters can be explored within this framework, some of which may provide further insight into the relationship between behavior and very specific brain region connectivity and others that may more accurately describe individual differences and state changes.

#### 4.1.1. Directed connectivity measures

Multivariate autoregressive (MVAR) approaches also allow for the parametric study of direct connections of a network while remaining immune to the influence of indirect connections on the network (Bastos and Schoffelen, 2015; Kus et al., 2004). Here, we describe the BRC results with a bivariate measure, WPLI, to study the indirect phase relationships between brain regions and behaviors. Using this bivariate approach, we are able to track the state of an individual based on their behavior and the intensity of bidirectional connections in the brain. In contrast, utilizing an MVAR-based connectivity metric such as the direct Directed Transfer Function (dDTF), allows for the study of direct unidirectional connections as they covary with a given behavior. A significant caveat to utilizing an MVAR approach lies in the likelihood of over-fitting such a model given limited information (i.e., a small window of trials). This becomes particularly problematic when modeling in sensor-space using a high-density headset (e.g. 64-channel or more) which leads to an ill-conditioned system that needs to be solved. This limitation may be resolved once in source space with sources consolidated into regions of interest (ROIs).

#### 4.1.2. Source space solutions

Analyzing all sensors from a spatially-dense system (e.g., a 256-channel vs. a 64-channel system) becomes computationally prohibitive as the number of connections increases substantially (i.e., 32640 sensor pairs compared to 2016). Conversely, analyzing sources allows for an extension of this approach by reducing the number of channels to a select number of ROIs while also potentially reducing noise and volume conduction. However, solving for the inverse problem presents a challenge as it is ill-posed and results can vary drastically depending on the assumptions implemented in approximating a solution. Within the context of small numbers of ROI source-space solutions, the use of an MVAR model may be applied on either individual trials or a group of contiguous trials. Additionally, the use of a blind-source separation technique (BSS) such as ICA or functional source separation (FSS) (Porcaro et al., 2009, 2013) provide an alternative means to explore the brain to connectivity relationship utilizing a consolidated distribution of sources and thus, connections. While single-trial estimates of WPLI and other connectivity estimates may be obtained using parametric approaches, such as MVAR models, the likelihood of overfitting such a model in sensor-space when using a high-density headset (e.g. 64-channel or more) is very high. In our experience, the parametric approach has been more problematic than beneficial when limited data is available, as is the case with the BRC approach.

### 4.1.3. Frequency range

The frequency band of interest is largely dependent upon both the nature of the task and the anticipated state(s) involved. For a target identification task, for example, several studies have focused on the alpha frequency band as it plays a role in attention (e.g., Thut et al., 2006; Busch et al., 2009) such that decreased phase and increased alpha activity are separately associated with decreased performance. While the alpha frequency band was investigated across both tasks in this paper to establish uniformity across example datasets, any frequency band may be chosen. The frequency band of interest should be selected on the basis of its studied relationship to the task and/or behavior of interest, such as the theta band to study working memory load (Shou and Ding, 2013) or the beta band for the study of language (Passaro et al., 2011).

### 4.1.4. Time period

The time-period of interest will also allow for some diversity when utilizing this approach. Relationships between connectivity and behavior may be studied in either a temporally congruent manner (track the connectivity time period to the occurrence of the behavior) or a predictive manner (track the connectivity time period to some point before the behavior occurs). Furthermore, the temporal length of the shifting window may be modified as necessary to emphasize a particular state or behavior, using one of the four window weighting functions examined here or an alternative one. The time duration of each epoch included within a window may also be modified. The frequency of interest will determine the minimum length of the window in most cases; for example, at least 1 s of data must be used to compute the connectivity for a 1 Hz period. Maximum trial lengths are dependent upon the interval between task-relevant events to ensure no overlap exists from one event to the next. Within the task timing and frequency constraints, the amount of data included for each trial is largely dependent on the preference of the investigator and is akin to preferences one has for determining pre-stimulus baseline periods (e.g., Luck, 2014).

### 4.1.5. Stability of BRC estimate

Different window lengths of trials and weighting functions may be applied depending on the stability of the investigated behavior and to place an emphasis on earlier or later trials. In studying stable representations of brain states that change slowly over time, such as fatigue levels during a monotonous task like interstate driving, a longer time window of trials/time may be utilized (e.g., windows of 150 1-s trials). Conversely, tasks and their associated behaviors that reflect transient states, such as vigilance during a target-detection task, may require shorter time windows (e.g., 10 1-s trial windows) to properly capture the associated neural representation. State fluctuations are difficult to control with high certainty within an experiment and critical parameters such as window length of trials partially alleviate this concern by identifying a robust explanation of the observed brain to behavior relationship regardless of the individual's state changes. The upper limit of window lengths of trials is dependent on the total time of the task as too many trials (e.g., 100-trial windows for a 150-trial task) reduces the degrees of freedom within the regression significantly along with a host of other problems concerning the computation of the phase-based connectivity. The weighting applied to each window may vary not only in its underlying function (exponential, logarithmic, Hanning, boxcar), but in direction as well, by placing an emphasis on subsets of trials within the window depending on the way in which the connectivity covaries with behavior.

When examining the effect of the number of trials and the weighting functions on the results, a visual verification identified an asymptote of statistically significant connections for a point when the variability was less than 10% change from one window length of trials to the next. The exponential weighting provided the least emphasis on trials before the last trial within the window of trials and was shown to asymptote sooner than the other three weighting functions tested. Since the behavioral fluctuations were identical and the connectivity values were different across all 4 weightings, this suggests that perhaps the BRC relationship is not driven only by how connectivity and behavior covary but also by the stability of the connectivity estimate. This result indicates that fewer trials within a window leads to less consistent BRC results. Consequently, the effect of window size should be investigated when applying the BRC approach to a novel experimental paradigm.

### 4.1.6. Physiological measures

The primary emphasis of the approach described here is on the behavior and its relationship to dynamic changes of the brain as observed through EEG. In theory, the behavioral metric could be replaced in this method with a physiological measure such as heart rate or blink rate. Utilizing this approach would allow for the study of not only a non-task-related component as it covaries with connectivity, but also a means by which the variance observed in the EEG signal could be explained by non-behavior sources. Typically, peripheral measures of this activity are viewed as noise in the EEG signal and subtracted out from the original signal, however, by utilizing the BRC approach that focuses on one or several of these other physiological measures, it might be possible to tease apart those connections which only appear to covary with non-neural sources.

### 4.2. Interpretation and considerations

The BRC results across both tasks here provide a complimentary understanding of the neural correlates of tasks and associated behaviors. For example, the networks identified by the conventional WPLI analysis appear to have little overlap with the connections from the positive and negative BRC analyses, suggesting that the connections derived from the conventional connectivity analysis across all trials are not directly related to fluctuations in these behaviors. This highlights the value of BRC since it directly examines time-varying behavior to better delineate the connections that are associated with both increases and decreases in performance. Another example of the complimentary component that a BRC analysis provides is demonstrated within the finger-tapping task, for which a commonly reported sensorimotor network was identified (e.g., Muthukumaraswamy et al., 2004). The BRC analysis yielded a different network of connections over sensors that do not seem to overlap with a sensorimotor network, and furthermore, they only appear to provide a relationship with behavior if the EEG activity of interest occurred during a period after the button had been pressed. This was in contrast to the target identification task, which highlighted a large network of connections both before and after the action of interest (i.e., target stimulus onset).

An important consideration lies in the assumption that an individual is likely to experience various cognitive states throughout the duration of an experiment. For example, a participant may be more vigilant at the beginning of an experiment and demonstrate a specific relationship between their behavior and the underlying task-specific connectivity pattern, however, as the task progresses, that individual may experience increased levels of mental fatigue leading to a change in both the underlying task-relevant network and the overall behavior. Testing for parameter convergence aids in establishing a robust approach to study the brain-behavior relationship in the face of state changes. Additionally, due to state fluctuations, it is unlikely that the same individual would produce the same BRC pattern of connections unless a near-identical state was observed. A large cohort of participants are needed to properly identify different cognitive states across retested individuals performing the same task such that state-unique BRC relationships would emerge as clusters of similar patterns.

An extension of the BRC approach outlined here could allow for a focus on the behaviorally predictive nature of the analysis within the context of a BCI. The past decade has seen a substantial influx of BCI-related EEG publications and applications focusing on behavior prediction (e.g., Trejo et al., 2005; Hammon et al., 2008; Wan and Makeig, 2009; Park et al., 2014; Touryan et al., 2016). A substantial technical challenge in this research has been identifying robust features to predict behavior. Shifting the focus of the BRC analysis to either the connectivity of the immediate time period preceding behavior (Figs. 5 & 6), or the connectivity in one or more trials before the behavior occurs, demonstrates a predictive relationship with behavior. Moreover, rather than focusing on many thousands of connections, the BRC approach can be modified and simplified to focus on an aggregate graph theory measure such as assortativity, betweenness, degree, etc. (Bullmore and Sporns, 2009). This modification would provide an alternative characterization of the behavioral fluctuations within the context of networks. Additionally, this allows for a single value across all sensors to track with behavior providing a more extensive parameter search across other components such as frequency, time, and trial delay.

### 4.3. Potential pitfalls

There are several caveats to applying the BRC approach across a wide range of tasks and behaviors without *a priori* knowledge of the expected direct brain-to-behavior relationship, the stability of the underlying neural response, and the variability inherent to the behavior of interest. For example, the BRC approach may be applied to a real-world driving task with a behavioral focus on blink-rate. In this situation, the behavior of interest may change rapidly and unexpectedly depending on the environment, and as such, the underlying neural correlates may not track well with sudden changes of behavior if they are studied across a broad window of time. One modification of the BRC method to address more transient fluctuations in behavior may involve computing bivariate phase-based connectivity measures on single trials by leveraging the short time period preceding the behavior ($-8$ s before, for example) and applying a sliding window across time-points within that time period.

## 5. Conclusion

The BRC approach allows for the study of connectivity as it covaries with behavior to complement the knowledge of underlying neural activity derived from conventional connectivity analyses. Within this framework, this method can be used to understand individual variability that is often treated as noise in most group-level neuroimaging studies rather than an informative component about behavioral performance. This analysis design also allows for an implementation within a BCI framework in an attempt to predict behavior in a trial-specific manner for a single subject. Consequently, applications of this methodology may provide further insight into the way in which neuroimaging data directly relates to a wide range of behavioral fluctuations.

### Acknowledgements

Research was sponsored by the U.S. Army Research Laboratory, including work under Cooperative Agreement Numbers W911NF-10-2-0022. The views and conclusions contained in this document are those of the authors and should not be interpreted as representing the official policies, either expressed or implied, of the Army Research Laboratory or the U.S. Government. The U.S. Government is authorized to reproduce and distribute reprints for Government purposes notwithstanding any copyright notation herein.